\documentclass[referee]{aa} 
\usepackage[varg]{txfonts} 
\usepackage{natbib}
\usepackage{graphicx}
\usepackage[utf8x]{inputenc}
\begin{document}

\title{On The colour of the Dark Side of the Moon\thanks{Accepted for publication in Astronomy\&Astrophysics Jan 8 2014.}}
\author{P. Thejll\inst{1}
\and 
C. Flynn\inst{2}
\and  
H. Gleisner\inst{1}
\and 
T. Andersen\inst{3}
\and 
A. Ulla\inst{4}
\and
M. O-Petersen\inst{3}
\and 
A. Darudi\inst{5}
\and 
H. Schwarz\inst{6}
}
\institute{Danish Climate Centre at DMI, Lyngbyvej 100, 
DK-2100 Copenhagen \O, Denmark
\and
Centre for Astrophysics and Supercomputing, Swinburne University, 
Hawthorn VIC, 3122, Australia
\and
Lund Observatory, Lund, Sweden
\and
Departamento de Física Aplicada, Universidade de Vigo, 
Campus Lagoas-Marcosende s/n, 36310, Vigo, Spain
\and 
Department of Physics, Faculty of Science, University of Zanjan, Zanjan, Iran.
\and
Leiden Observatory, NL-2333 CA Leiden, The Netherlands}

\abstract {} 
{``Earthshine'' is the dim light seen on the otherwise dark side of the Moon,
  particularly when it is close to new. ``Earthlight'', or reflected sunlight
  from the Earth, is the source of Earthshine. 
  Using $B$ and $V$ band CCD images
of both the dark and bright side of the Moon, we aim to estimate the Johnson photometry $B-V$ colour of the
Earthshine for the first time since the late 1960s. From these measurements we are also
able to quantify the colour of Earthlight.}
{We present images of the Moon taken with a small refractor in Hawaii, in $B$
  and $V$ bands and taken under favourable conditions so that
  scattered light in both bands almost completely cancels, yielding a
  map of the surface in $B-V$ colour. Co-addition of 100 such images taken in
  rapid succession substantially improves the signal to noise ratio, and several
  sources of photometric bias are eliminated by use of relative methods.}
{The earthlit dark side of the Moon is observed to be $0.150 \pm 0.005$  mag
  bluer in $B-V$ than the sunlit bright side, in good agreement with the only
  known previous measurement of this quantity from 1967. Arguing on the basis of the change in
  $B-V$ for sunlight reflected once off the Moon, we derive a colour for
  earthlight of $B-V = 0.44\pm0.02$ mag (without applying a small, uncertain,
  phase-dependent reddening correction).  The absence of a colour-gradient
  in the $B-V$ image implies that the scattering properties of the
  atmosphere+optical system are almost exactly matched in the two
  wavelength bands, the consequences of which are discussed.} {}

\titlerunning{The colour of the dark side of the Moon}
\authorrunning{Thejll et al.}
\maketitle

\section{Introduction}

The study of the photometric properties of the Lunar surface is in a phase of rapid
development through both ground-based ~\citep[e.g.][]{2011Icar..214...30V}
and space-borne ~\citep[e.g.][]{2011Icar..215..639Y,2012LPI....43.2810Y,
  2012LPI....43.2795B} instruments. Absolute photometry of the sunlit Moon
(hereafter BS for the ``Bright Side'') can be used to calibrate
satellite borne instruments -- notably Earth-observing
satellites~\citep{2008spie.7081e..28s}, and is an important means of
ensuring long-term stability in data products, such as used for climate
studies.

The so-called ``Dark Side'' of the Moon (hereafter ``DS'') is almost
always illuminated by some fraction of the Sun-lit Earth. This light
is called ``Earthshine'', and is to be distinguished from its source,
which we hereafter refer to as ``Earth{\it light}''.

Photometry of the bright and dark sides of the Moon can be used
to measure/monitor the changing albedo of the Earth. Pioneering
photometer-based work in this field was undertaken already in the
1930s~\citep{1933anost...3..138d,1943c&t....59..375d}, and also later
~\citep{1964saosr.162.....b,1966ArA.....4..131R}.  Modern studies,
utilizing CCDs, date from the work of ~\cite{2001georl..28.1671g}, who show
that albedo studies of the Earth can be performed from the Earth, with
accuracies rivalling satellites and precisions limited mainly by the
Earth's natural albedo variability. A global network of about half a
dozen small telescopes could in principle monitor
Earth's changing albedo via Earthshine photometry rather cheaply, and
with much better long term stability, than satellites ~\citep{flatteetal}.

Satellite measurements of visual-band fluxes from the CERES and ERBE
instruments have shown \citep{2006tella..58..320b} that the global-mean
annual cycle in monthly-mean albedo has an amplitude of about 3\% of the
mean value.  Shorter-term variations can be larger -- in GERB full-disc
images at 15 minute intervals, it was found ~\citep{Oelund2013} that
at quarter phase albedo varied by 20\%, due to changing cloud masses.

Earthshine telescopes currently in use are strongly motivated by the design
made at Big Bear Solar Observatory
~\citep{2001georl..28.1671g,2010adast2010e..26g} and collaborators. Telescopes
of this type are currently operating from Tenerife and the Crimea. The
instruments are small (few cm aperture) automatically operating refractors,
which image the Moon and the surrounding sky on a suitable scale (near $1^\circ
\times 1^\circ$).

The essential observational issues to be solved are, firstly, the reduction
and/or removal of the significant light scattering, arising partly in the
atmosphere, but mainly in the optical train of the telescope/detector; and
secondly, the high contrast ratio between the BS and DS, which typically is not
much less than the dynamic range of the imaging (CCD) device.

The telescopes typically consist of a primary lens and secondary optics
designed to allow the insertion of a Lyot-stop, to eliminate divergent
light rays~\citep{1939MNRAS..99..580L,1973JOSA...63.1399N}, along with a means of inserting
optical band- or neutral density-filters, and a shutter at the pupil. They
also typically use high quality bi-element objectives (in order to limit
the number of optical surfaces) and occulting devices near the prime
focus with which to cover the BS. A typical occulting
device is a knife-edge that can be manually or automatically positioned
in the field of view. Such devices suppress the light entering the
rest of the optical system, at the cost of producing a bright field of
forward-scattered light within the telescope.

Since such a device adds complexity to operational procedures, while not
altogether eliminating the need for subsequent numerical modelling of the
scattered light, we have considered doing without it. Numerical modelling of
scattered light may be unavoidable in the bulk of earthshine observations, but
attention to optical design, internal baffling and identifying excellent
observing conditions can also give interesting results. We find ourselves with
data from a rare and special night where the data show unique properties, and
discuss these here. 

We have operated our own equatorially mounted earthshine
refractor~\citep{2008SPIE.7012E..91O,2010spie.7733e..88d}, from
National Oceanic and Atmospheric Administration's observatory on Mauna
Loa (3397 m altitude) in Hawaii, during 2011 and 2012, acquiring 500+
image sequences of the Moon in 5 broad-band filters -- amongst them
Johnson $B$ and $V$.  Our imager is an Andor DU897-BV camera based on
a sensitive back-illuminated thinned CCD chip (512$\times$512 pixels,
pixel size of 16 $\mu$m and image scale 6.67 arcsec per pixel, field
of view is just under 1$\times$1 degree of arc). The detector is cooled
thermostatically by a Peltier element.

In this paper, we present $B$ and $V$ band imaging data of the Moon taken with
our Earthshine telescope, under unique atmospheric conditions in which the
scattered halos in $B$ and $V$ light cancelled so well as to yield a $B-V$
``colour map'' of the Lunar surface with virtually no residual scattered-light
colour gradient. This map has been used to measure the colour of Earthshine on
the DS, and sunlight on the BS, as a means of estimating the
colour of Earthlight.

We believe this to be the only recent determination of the dark side's
Johnson $B-V$ colour, since the late 1960s.

\section[]{Observations and Data reductions}

The waning crescent Moon was observed on JD2455945.177 (16:00 UTC,
January 18, 2012) at an airmass of $z~\sim$1.5 with an illuminated
fraction of close to 0.25; lunar phase near 40$^\circ$. Image stacks
of 100 consecutive 55 millisecond images in $V$ and 155 milliseconds
in $B$, along with bias frames immediately before and after the stacks,
were exposed.
Flat field images in $B$ and $V$ were acquired from a
Hohlraum lamp source mounted on the dome wall. The flat fields originally
had some smooth spatial gradients near the edge which were removed by
subtraction of a fitted low-order polynomial surface after which the
mean level was added back. Hundreds of images had been obtained in both
bands and averaged.  Flat field noise is thus low and the flat fields
are dominated by dust speck signatures and a fixed diagonal pattern
probably due to CCD thinning.

The bias level in the CCD oscillates by about 0.25\% over a 20-minute
period due to the Peltier cooling element. To remove this, we create a
'superbias' image from hundreds of bias images that were averaged,
using the mean half-median technique on each pixel, to yield a very low-noise
image of the bias. Bias frames taken before and after science frames could then
be used to scale the superbias to the appropriate level. This procedure proved
very satisfactory. The science frames were bias-subtracted, flat fielded
following standard techniques, and then converted to fluxes by dividing by the
exposure times.

Although atmospheric extinction coefficients were not measured on the night of
the observations, long term monitoring of extinction during the project has
yielded values of $k_B=0.17$ and $k_V=0.08$. These coefficients are needed to
set the zero point of the photometry, but have no effect on the colour
difference of the DS and BS, since they factor out.

Transformations from instrumental to Johnson $B$ and $V$ were derived
from observations of the open cluster NGC 6633 (RA = 18h 37m and DEC =
+06 34, J2000), for which colours and magnitudes are available in the
WEBDA Open Cluster database (http://www.univie.ac.at/webda/). This
cluster was very convenient, because (1) is ~20 arcmin across, ideal for
our camera's field of view. It is young ($<$1 Gyr), so there is a wide range of colours
amongst the stars, from $0 < B-V < 2.0$, while the stars are relatively
bright ($V<12$) and can be reached in less than a minute in $B$ and $V$).  Photometry in $UBV$ is available in ~\cite{1958apj...127..539h} for 95 stars with $V<12$, and these stars were located in our
images and transformations obtained to the standard Johnson system. The
estimated accuracy of the $V$ magnitudes in ~\cite{1958apj...127..539h} is $\pm
0.012$ mag and in $B-V$ colour is is $\pm 0.006$ mag. Averaging over
95 stars leads to a very precise calibration.
The scatter of the transformations is about 0.02 mag, while the
zero-point is set to better than 0.01 mag.

The lunar images were carefully aligned, as miss-alignment tends to
leave surface-feature patterns and residual light of the two halos
(i.e. scattered light) in $B$ and $V$ in the colour map. Optimal
alignment was found by using a combination of visual inspection and
correlation methods.  Typical standard deviation of the alignment
errors are at the sub-pixel level; worst-case misalignments above half
a pixel are easily caught by visual inspection. Given the plate-scale,
we estimate alignment errors to be 1-2 arc seconds generally, 3 arc
seconds at worst. In terms of degrees of lunar longitude these numbers
correspond to 0.2$^\circ$, at most. The disadvantage of using a method
based on co-adding 100 aligned images is that the signal to noise ratio
(SNR) is not quite as high as that available using an occulting device --
we expect a single well-exposed image taken with such a device covering the bright
side to have a DS SNR 5 times higher than what is possible with 100 co-added
short exposures. Alignment of images is possible to a  satisfactory degree,
but the limited resolving power of our optical system spreads the light
from e.g. bright lunar highlands into darker mare, making use of the
co-located selenographic coordinate system, and choice of lunar regions
for photometry, critical. The disadvantages could be countered by using
a faster optical system with better resolving power and a camera with a
larger buffer and faster image download speed. The advantage of the co-addition
technique is that a symmetric ray path is maintained so that an axially
asymmetric scattered light field inside the telescope/camera system is
avoided. 

Colours were measured on the dark and bright sides at particular regions
of the surface, which were extracted from the pixel space images on the
basis of a selenographic coordinates. The mapping between image pixel
positions and the selenographic coordinate system is possible with the
ephemerides and some geometry. We selected regions centered on those
used by ~\cite{franklin1967}: that is, Mare Crisium at 18$^\circ$ N,
60$^\circ$ E, and Oceanus Procelarum at 19$^\circ$ N, and 60$^\circ$
W. Photometry was performed at these coordinate centers for rectangular
apertures of sizes from plus/minus 1 to plus/minus 5 selenographic
degrees (in longitude and latitude).  We determined an optimal size
of plus/minus 3 degrees. Larger apertures tended to erroneously
include a few pixels from adjoining, higher or lower luminosity lunar areas;
smaller apertures contained fewer pixels and noise averaging suffered.

The mean values in each subregion, and the standard deviation of
that mean were obtained by bootstrap sampling (5000 times) with
replacement~\citep{efrontibshirani93} on the pixels contained in the
sub-images.

\begin{figure}
\includegraphics[scale=.45,viewport=40 0 556 541,clip]{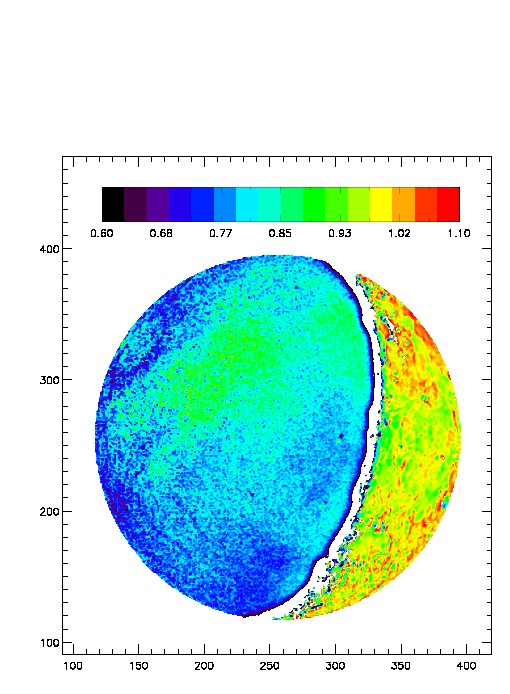}
\includegraphics[scale=.45,viewport=40 0 556 541,clip]{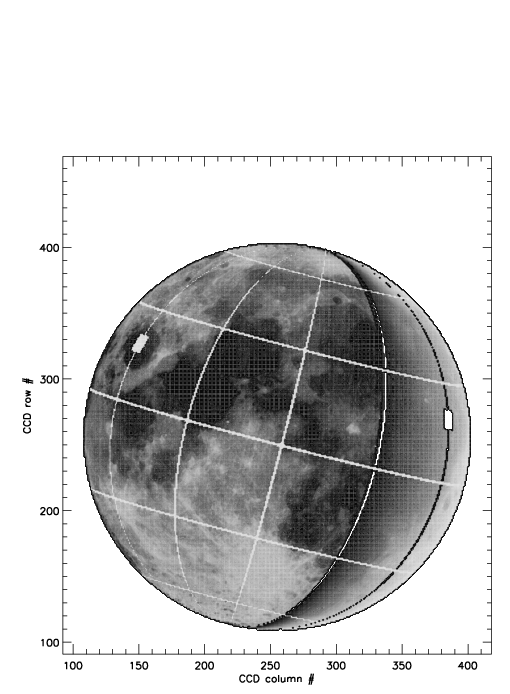}
\caption{$B-V$ image (upper panel) of the waning crescent Moon on
JD 2455945.177. False colours represent $B-V$ values; the DS is blue
and to the left in the figure; the bar shows $B-V$ and corresponding
colours. The finder chart (lower panel) shows the measurement areas
in Mare Crisium and Oceanus Procelarum. Subtle colour differences,
corresponding to M. Crisium, M. Fecunditatis,  M. Nectaris, Tycho and
adjoining M. Nubium can be recognized using the finder chart. Lines of
equal longitude and latitude are drawn at 30$^\circ$ intervals. Eastern
longitudes are to the left in the images. }\label{fig:fig1} \end{figure}

\begin{figure}
\includegraphics[scale=.50,viewport=11 229 553 795,clip]{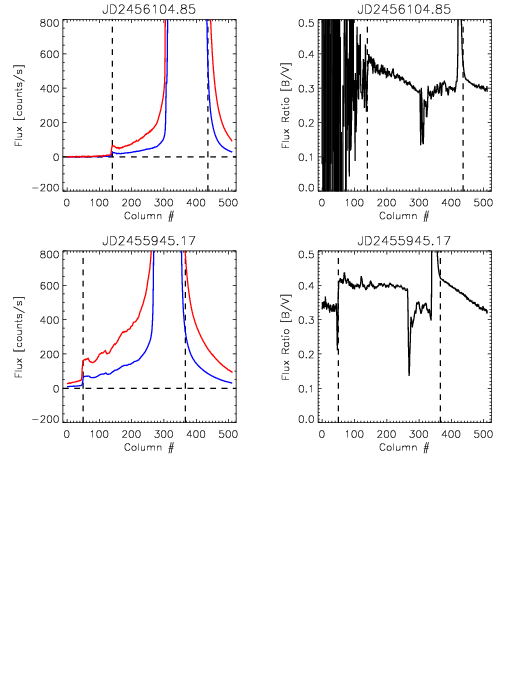}
\caption{Row-slices across the Moon, through disc centre, on two nights,
to illustrate halo profiles and intensities. The upper row is from
a night where the $B$ and $V$ halos did not cancel, while the lower
row is from the unique night JD2455945.17, where the halos did cancel
on the DS part of the lunar disc. The left column shows profiles (B
is in blue, and V is in red), while the right column shows the $B/V$
flux ratio. Horizontal dashed line shows the zero flux level. Vertical
dashed lines show the disc edges. Orientation of DS and BS are as in
Figure~\ref{fig:fig1}.}\label{fig:fig11} \end{figure}

A calibrated false-colour $B-V$ image of the Moon is shown in
Figure~\ref{fig:fig1}. There is almost complete cancellation
between the halos of BS-scattered light in the $B$ and the $V$ bands.
In Figure~\ref{fig:fig11} details of the halo profiles on this night,
and a night where halos do not cancel, are shown. The night JD2455945.17
is mediocre in terms of the strength of scattering in the
atmosphere -- the other night shown is much clearer, but on that night
the halos do not cancel. The cancellation is an as-yet poorly understood
interplay of halo profiles and intensities dictated by lunar phase and
atmospheric conditions.

In Table~\ref{tab:tab1} we give calibrated magnitudes from selected
regions (see caption of table for details) in the $B$ and $V$ images
seperately, along with $B-V$ calculated from the same regions in the $B-V$
image directly. We see that the difference between $B$ and $V$ magnitudes
is the same as $B-V$ found in the $B-V$ image, but that the errors of the
former difference is larger than the latter. This constitues a
sanity check on the data. $B-V$ for the same regions extracted in a
Full Moon image from JD2455814 were also extracted so that the intrinsic
surface colour-differences between the two regions could be estimated ---
results likewise in Table~\ref{tab:tab1}.

The literature since the late 1960s contains discussions of how much
lunar colour depends on the phase angle; the dependence is low but can
be measured \citep{1969AJ.....74..273M,velikodsky2011}. Observations,
from the ground and from lunar orbiters, of lunar reddening in $B-V$
can be extrapolated to the relevant phase for our observation (DS near 0
degrees phase, BS at 40 degrees phase) and yields a suggested reddening
of at least 0.06 mags of the DS-to-BS observation. Observations of
the phase-dependent reddening of lunar colours, since it was first
noticed and until now, do not agree on the magnitude of the
effect, only the sign -- e.g. compare ~\cite{1969AJ.....74..273M}
and ~\cite{1973aj.....78..267l}. Since we wish to compare to
\cite{franklin1967}, who did not apply any phase-dependent corrections,
nor do we.  Franklin's set of observations cover phase angles from 32
to 78 degrees, spanning our single observation. There is no hint of any
phase dependence in $B-V$, but the scatter in Franklin's $B-V$ data is
of the same order of magnitude as the expected reddening. 

\begin{table}
\caption{Results. Average $B$ and $V$ and average $B-V$ difference, and
uncertainties, for two selected regions on the Moon.  All magnitudes are
absolute calibrated, using stars in the open cluster NGC 6633 (see text
for details).  B image was obtained at JD 2455945.1760145 while V image
was obtained at JD 2455945.1776847.  Means and uncertainties found using
bootstrap sampling with replacement on sub-image region pixels. The
BS region lies in Oceanus Procelarum (OC), centered at 19$^\circ$
N and 60$^\circ$ W, while the DS region is in Mare Crisium (MC) and
is centered at 18$^\circ$ N, 60$^\circ$ E.  We report here results for
box-size of width plus/minus 3 selenographic degrees.  The third line
gives $B-V$ extracted from the DS and BS regions of the $B-V$ image
itself while lines 1 and 2 are based on extractions from the $B$ and $V$
images individually. Note that $B$ and $V$ magnitudes are dependent on
knowing the extinction corrections while $B-V$ is not. 'DS' and 'BS'
refer to the observations on JD2455945, while 'FM' (lines 4 and 5)
refers to the near-Full Moon observations of JD 2455814. No corrections
for phase-dependent reddening have been applied.} \label{tab:tab1}
\centering
 \begin{tabular}{rlrrr}\hline\hline
 & Location   & $\left<{B}\right>$ & $\left<{V}\right>$ & $\left<{B-V}\right>$ \\\hline
 1& MC (DS) & 10.166$\pm$0.009&9.420$\pm$0.006&0.75$\pm$0.01\\
 2& OP (BS) &  2.093$\pm$0.006&1.194$\pm$0.007&0.90$\pm$0.01\\

\cline{1-5}
 3& DS-BS   & -      &     -         &$-0.150 \pm$0.005\\

\cline{1-5}
 4& MC (FM)       & - & - & 0.883$\pm$0.001 \\
 5& OP (FM) & - & - & 0.832$\pm$0.002 \\  \hline
\end{tabular}
\end{table}

The colour of the Sun has been determined indirectly using Sun-like
stars by \citep{2006mnras.367..449h}, who obtain $(B-V)_\odot = 0.64 \pm
0.02$ mag. Another method to determine the colours of the Sun is by
applying synthetic photometry to the observed, absolute flux calibrated
solar spectrum; this results in $(B-V)_\odot = 0.63 − 0.65$ ~\citep{1996AJ....112..307C,1998A&A...333..231B}, in good agreement with
the \cite{2006mnras.367..449h} value. For the BS (which is illuminated directly
by the Sun), we derive a colour of $0.90 \pm 0.01$ mag: thus, the Moon
reddens the Solar spectrum by $0.90 - 0.64 = 0.26$ mag, on average. If
we assume the same reddening of the Earthlight when it reflects from
the Moon as Earthshine, we derive an apparent colour for Earthlight on
this particular night of $B-V = 0.75 - 0.26 = 0.49 \pm 0.02$ mag. The
uncertainty given is dominated by the uncertainty on $(B-V)_\odot$.

However, we have also estimated, using Full Moon images from JD 2455814,
the intrinsic difference in $B-V$ colour between the selected areas
in Mare Crisium and Oceanus Procelarum to be 0.051$\pm$0.002 mag with
Mare Crisium being redder. This implies that the {\it true} colour of
the Earthlight, as seen from the Moon, was 0.051 mag bluer than the
Earthshine, or that the $B-V$ for Earth{\it light} is $0.49 - 0.051 =
0.44\pm0.02$ mag.  Since phase-dependent reddening occurs
when light strikes the Moon the value is a lower limit on earthlight's $B-V$
colour. The phase-dependence of the reddening is poorly known so
we do not correct for it -- but orders of magnitude can be given:
~\cite{1969AJ.....74..273M} measured (at 90 degree phase, for reference)
a 3\% effect, while ~\cite{1973aj.....78..267l} found a 10\% effect
at the same angle, so the order of magnitude of the reddening for the
relevant observations we, and Franklin have are from several hundredths
of a magnitude to tenths.

A strong scattered-light halo is typically present surrounding the BS in
any image of the Moon. Typical $B-V$ images obtained at other phases and
other nights show strong spatial gradients.  Both the $B$ and $V$ images
we use here show halos (Figure~\ref{fig:fig11}), so the absence of a
gradient in the $B-V$ image implies conditions such that halos cancelled.
By selecting a crescent phase, and undoubtedly being lucky with conditions
on the given night we have found halo strengths and profiles in $B$ and
$V$ that are matched.  This may have implications for understanding the
origin of the scattering. If the halos are due to wavelength-dependent
processes --- such as atmospheric Rayleigh-, and to a lesser extent
Mie-scattering, then they would not be expected to cancel so cleanly ---
and this is certainly the case on most other nights. Scattering also
originates in the telescope optics from micro-scratches and optical
imperfections, and aberration due to the optics. The Lyot stop removes
most of the non-axially scattered light that enters the first of the
secondary lenses, but a remnant is left which produces at least part of
the halo. This is seen by inserting the occulting device in the prime
focus and imaging various sources --- behind the solid knife edge a halo
is evident; thus at least part of the halo is formed in optics or devices
that follow the prime focus. On less perfect nights a contribution to the
halo is also made by the atmosphere --- it is stronger and also wavelength
dependent and thus cannot be expected to be removed by subtraction.
This argument implies that whatever processes are causing scattering in
the telescope itself are either not wavelength dependent or are very weak.

\subsection{Error limits of the technique}
We can assess the minimum possible error in the $B-V$ determinations by
using synthetic Lunar models with appropriate Poisson noise, but with
all problems due to realistic observing limitations omitted. We have
a code which can produce artificial images of the Moon for any of our
observations from Hawaii, and we use bootstrapping with replacement
on subregion pixels on an appropriate artificial image and repeatedly
evaluate the mean colour difference. Extracting the standard deviation for
the distribution we can determine the uncertainty due to Poisson noise,
the centering of the moon, and pixel selection. We find that errors can
be as low as $\pm$0.002 mag ideally (whereas we observe the error to be
$\pm$ 0.005 mag) in comparisons between a DS and BS region. The minimum
error when comparing two BS regions is $\pm$0.0004 mag compared to about
$\pm$0.002 mag in the case of the real observed images.  We conclude that
there is a factor of from two to five between the lowest errors possible
given Poisson statistics and what we find in real data.  Improvements
could potentially be made in our analysis, particularly in areas related
to bias subtraction, flat fielding, and image alignment. Reduction in
errors beyond that could be based on using more than 100 images per
stack, which is the current limit of the image-buffer in our camera.

\section[]{Comparison to other work}

The literature on lunar dark-side photometry and colorimetry is dominated by
the early works of Danjon and Dubois
\citep{1933anost...3..138d,1943c&t....59..375d,1947BuAst..13..193D,1955lastr..69..242d},
but unfortunately their pioneering work is difficult to interpret in terms of
modern photometry. This is quite a pity, because they have produced the longest
set of Earthshine colour measurements. In trying to make use of their
photometry, we are mainly hampered by a lack of understanding of the
light-scattering properties of their equipment, and of the atmospheric
(scattering) conditions on individual observing nights. Furthermore, their
observing procedures did not compensate for the spatial gradient in the
scattered light.  However, if we use the decadal-mean values given by
\cite{1933anost...3..138d} -- and note the addendum provided there --
comparison with our own work is possible.  Danjon worked in the ``Rougier''
filter system and gives colour index values for earthshine as well as
earthlight. The data can be converted to Johnson $B-V$ by use of the Rougier
colour index for the sunlit Moon, the Sun and their known $B-V$ values
\citep{1964jgr....69.4661w}.  We find that $B-V$ corresponding to Danjon's
average data is 0.35 with limits from 0.12 to 0.53. The wide limits are stated
by Danjon as being due to seasonal variations in earthshine colour as seen from
their observing site in France.

The only other measurement of Lunar dark-side in Johnson photometry which we
could find in the refereed literature is by \cite{franklin1967}, who performed
DS aperture photometry referenced to standard stars and reported a difference
between $B-V$ from the DS to the BS of $-0.17$, on average (i.e. DS bluer) for
Mare Crisium and Ocean Procelarum locations.  We have used his data to
calculate the standard deviation in this value and find $\sigma = 0.05$ mag.
In this paper, we find a colour change of $\Delta(B-V)_{DS-BS}= -0.150 \pm
0.005$ mag from our observations, so the two
measurements are consistent.

Recently, the Cassini Solstice mission acquired images of the Earth and Moon as
seen from beyond Saturn --- unfortunately, pixels in the image of Earth (in the
press-release images) are saturated and photometric analysis of these data does
not appear possible. Potentially, the original data could be utilized, and we
are continuing work on this. A similar situation applies to recent Juno
spacecraft images.

Instruments in Lunar orbit can observe the DS, but it is still challenging to
do so as the Earthshine is intrinsically faint and exposure times become so
long that image blur can occur due to platform motion.  The Lunar
Reconnaissance Orbiter has taken several images of the DS, available in the Jet
Propulsion Laboratory public data archives, but inspection shows that only very
faint DS levels are present because the exposure time settings are set
optimally for imaging of the BS.

Low-Earth orbit satellites daily take images of Earth, and in many
spectral bands, but we have found no conversion to the Johnson system. LEO
imagery is unsuited for building instantaneous full-disc images of
Earth, due to the long revisit time for a given area.  Geo-stationary
satellites take full-disc images, but only in one visual and a number
of near-IR and longer bands, which does not allow for a transformation
to the Johnson system.  No readily useful spacecraft-based Johnson $B-V$
results for the lunar DS or the earthlight therefore exist.

Observed spectra of Earthshine have
been published~\citep{
1936PAAS....8R...7D,
2002ApJ...574..430W,
2002A&A...392..231A,
2005ApJ...629.1175M,
2006ApJ...644..551T,
2006A&A...460..617H}
but are mainly unsuitable for spectro-photometry in the visual wavelength
range as they are presented mainly for discussions of telluric
signatures in the red, near-IR and IR parts of the spectrum, and
not  in terms of absolute flux calibration or photometric colours. We
contribute calibrated Johnson $B-V$ colours as a first step in this
direction. Other papers are devoted to polarimetry of the earthshine
~\citep{2013A&A...556A.117B,2013PASJ...65...38T}, and their results are
not suited for spectro-photometry. \cite{2013MNRAS.435.2574G} show
normalized high-resolution spectra in a narrow wavelength range
only.  On the basis of the earthlight spectrum for May 31 2005,
in ~\cite{2006A&A...460..617H}, we estimate $(B-V)_{EL}=0.53\pm0.05$,
which is redder than what we derive. Daily variations in the earthshine
spectrum shown in ~\cite{2006A&A...460..617H} correspond to 0.05 mags
in $B-V$. Our own estimates of daily variations of earthshine flux, based
on satellite data, is 20\%, at quarter phase, so we see no disagreement
between our observation and the one based on published spectra.

Of related interest to us are the GOME (Global Ozone
Monitoring Experiment) disk-averaged observations of the (Sun-lit)
Moon~\citep{1997ESASP.414..743D,1998ApOpt..37.7832D} --- the results for
the wavelength dependent Lunar albedo can be used to determine BS $B-V$,
based on reference spectra of the Sun, such as the~\cite{Wehrli-1985}
spectrum. Multiplying this spectrum with the GOME albedo law,  which is
almost linear in wavelength -- Figure 4 in~\cite{1998ApOpt..37.7832D},
and applying $B$- and $V$-filter transmission curves, we derive
$(B-V)^{Wehrli,GOME}_{BS} = 0.88$ mag. This is 0.25 mag redder than the
Wehrli-based solar $(B-V)_{\odot}^{Wehrli}$ colour of 0.63 mag, which
we compute in the same way. The difference between observed values of
$B-V$ of the Sun~\citep[0.64$\pm$0.02 mag]{2006mnras.367..449h} and our
BS observations of the Moon is $0.26\pm0.02$ mag, consistent with the
GOME-based result.

We note that the colour for Earth given
in~\cite{2000allensqu.book.....C,1975allensqu.book.....A,1955QB461.Allen564.....}
is $B-V = 0.2$, but it is unclear whence the value originates.  In the
1955 edition of Allen, work by~\cite{1933anost...3..138d} is cited,
but as far as we can see, Danjon gives values for the colour index
inconsistent with $B-V = 0.2$.  Accompanying notes in Allen refer to a
reduction of the published value of the planetary colour index by 0.1
to 0.2 to satisfy observations, but the origin of this correction is
unclear. As seen above, the $B-V$ equivalent to the Rougier colour index
values places the Earthlight colour in a range consistent with the Allen
value of '0.2', but we suggest that the Danjon-based mean value of $B-V =
0.35$ mag be used instead.

\section{Summary}
We have obtained $B$- and $V$-band images of the Moon on JD
2455945.177. We construct a $B-V$ image and find area-mean values for the
dark-side to bright-side $B-V$ difference, using reference areas defined
in~\cite{franklin1967}. We have used relative photometric observing-
and data-reduction methods, thereby eliminating important sources of
potential bias such as dependencies on detector sensitivity variations
in time, extinction corrections and exposure times.  Currently, error
limits are not set by the inherent Poisson statistics, and errors could
potentially be lowered by factors of 2--5 with careful attention paid
to data-reduction techniques; after that improvements can be expected
by increasing the number of images used in the stacks.

We have shown that the $B-V$ colour of Earthshine, determined from a pair
of $B$ and $V$ image stacks, taken on a photometrically exceptional night
at Mauna Loa with our Earthshine telescope, agrees within errors with the
only other published dark side $B-V$ value, that by~\citet{franklin1967}
observed 45 years earlier, and is in general agreement with the broad
limits set by Danjon and Dubois' work 1930s-1950s.

This paper has been about a single pair of fortuitously identified
images. We have 500 more, but they are not suited for this type of
analysis --- their halos do not match. We cannot rely on the present method
for analysis of long-term evolution of earthshine unless a method can be
found to convert existing pairs of images to pairs where the halos match.
It may be possible to transform images suitably by adaptive convolution
methods.  Time will tell if that approach works; in the meantime work
will continue on a different type of analysis --- forward modelling of
the scattered light --- and albedos in 5 colour bands will be presented
in a future publication.

\begin{acknowledgements}
This work was supported by the Danish Climate Centre at The Danish
Meteorological Institute.  We thank VINNOVA for the funding that made
the earthshine project possible, and NOAA staff for their assistance at
Mauna Loa during observations.  Ana Ulla thanks the University of Vigo
for funding under grant 12VI20. M. Collados and R. López are thanked
for technical discussions on the optical configuration of the telescope.
Ben Berkey helped us extensively and knowledgeably at MLO and is warmly
thanked.  John Sarkissian is thanked for very useful discussions on Lunar
colours, and Luc Arnold provided access to his earthshine spectra, for
which we are most grateful. Comments and suggestions from an anonymous
referee are warmly acknowledged. This research has made use of the
WEBDA database, operated at the Department of Theoretical Physics and
Astrophysics of the Masaryk University.  \end{acknowledgements}

%%% References
\bibliographystyle{aa}
\bibliography{localrefs.bib}

\end{document}